\documentclass[aps,preprint,tightenlines]{revtex4}%
\usepackage{amsfonts}
\usepackage{amsmath}
\usepackage{amssymb}
\usepackage{graphicx}%
\providecommand{\U}[1]{\protect\rule{.1in}{.1in}}

\begin{document}
\title[Nanoshells as a high-pressure gauge analyzed to 200 GPa]{Nanoshells as a high-pressure gauge analyzed to 200 GPa}
\author{Nick Van den Broeck$^{1}$}
\email{nick.vandenbroeck@ua.ac.be}
\affiliation{$^{1}$Theory of Quantum and Complex Systems, Universiteit Antwerpen,
Universiteitsplein 1, 2610 Wilrijk, Belgium }
\author{Katrijn Putteneers$^{1}$}
\affiliation{$^{1}$Theory of Quantum and Complex Systems, Universiteit Antwerpen,
Universiteitsplein 1, 2610 Wilrijk, Belgium }
\author{Jacques Tempere$^{1,2}$}
\affiliation{$^{1}$Theory of Quantum and Complex Systems, Universiteit Antwerpen,
Universiteitsplein 1, 2610 Wilrijk, Belgium }
\altaffiliation{$^{2}$Lymann laboratory of physics, Harvard, 17 Oxford Street, Cambridge, MA
02138, USA}

\author{Isaac F. Silvera$^{2}$}
\affiliation{$^{2}$Lyman Laboratory of Physics, Harvard, 17 Oxford Street, Cambridge, MA
02138, USA}
\begin{abstract}
In this article we present calculations which indicate that nanoshells can be
used as a high pressure gauge in Diamond Anvil Cells (DACs). Nanoparticles
have important advantages in comparison with the currently used ruby
fluorescence gauge. Because of their small dimensions they can be spread
uniformly over a diamond surface without bridging between the two diamond
anvils. Furthermore their properties are measured by broad band optical
transmission spectroscopy leading to a very large signal-to-noise ratio even
in the multi-megabar pressure regime where ruby measurements become
challenging. Finally their resonant frequencies can be tuned to lie in a
convenient part of the visible spectrum accessible to CCD detectors.
Theoretical calculations for a nanoshell with a SiO$_{2}$ core and a golden
shell, using both the hybridization model and Mie theory, are presented here.
The calculations for the nanoshell in vacuum predict that nanoshells can
indeed have a measurable pressure-dependent optical response desirable for
gauges.However when the nanoshells are placed in commonly used DAC pressure
media, resonance peak positions as a function of pressure are no longer
single-valued and depend on the pressure media, rendering them impractical as
a pressure gauge. To overcome these problems an alternative nanoparticle is
studied: coating the nanoshell with an extra dielectric layer (SiO$_{2}$)
provides an easy way to shield the pressure gauge from the influence of the
medium, leaving the compression of the particle due to the pressure as the
main effect on the spectrum. We have analyzed the response to pressure up to
200 GPa. We conclude that a coated nanoshell could provide a new gauge for
high-pressure measurements that has advantages over current methods.

\end{abstract}
\startpage{1}
\maketitle

\section{Introduction}

Recently nanoshells\cite{NS} have received a lot of attention because they
possess interesting optical properties. Consisting of a dielectric core
surrounded by a metallic shell, nanoshells allow surface plasmon polaritons to
exist at the interfaces between the layers, resulting in a spectrum with very
pronounced absorption and extinction peaks. The position and broadness of
these resonance peaks depend on the size, shape and materials of the particles
and on the surrounding medium. This versatility and sensitivity make
nanoshells excellent probes in different fields, ranging from cancer
ablation\cite{cancer} and biosensors\cite{biosensor} to surface enhanced raman
spectroscopy applications\cite{sers} and plasmonics\cite{Plasmonics}. In this
paper we investigate the usefulness of nanoshells in the field of
high-pressure physics, and more precisely Diamond Anvil Cell (DAC) pressure
measurements\cite{DAC}. The first theoretical calculations for a spherically
symmetric SiO$_{2}$-gold nanoshell under pressure (up to 200 GPa) are
presented, indicating clear measurable absorption peaks. The main idea behind
the use of nanoshells as a pressure gauge is that the size of the nanoshell
and the permittivity of the SiO$_{2}$ core and medium will change under
pressure resulting in a shift of the resonance frequencies. The goal of this
paper is to calculate the optical transmission as a function of wavelength for
a spatial distribution of noninteracting monodisperse nanoshells.

At the moment there are several pressure gauges used in high-pressure physics.
We briefly discuss them and conclude with the advantages of the
nanoshell-gauge in comparison with the established methods. The prime and
venerable pressure gauge is the shift of the peak of the ruby fluorescence
spectrum with pressure\cite{Ruby,Rubyscale}, excited by laser pumping. In this
case micron sized ruby grains are embedded in the pressurization medium.
Several problems can arise above a certain pressure\cite{Ruby
error,Rubyto150GPa}: at pressures above 100 GPa diamond fluorescence can
become intense and mask the ruby line; ruby grains can be blown away during
loading or masked during pressurization; above a few hundred GPa the standard
method of pumping the ruby with a green or blue laser line becomes ineffectual
as the laser light is absorbed by the diamond anvil. Although some of these
problems can be overcome\cite{Eggert,Chen}, ruby becomes challenging to use at
very high pressure. For cases where the ruby is lost or cannot be observed
researchers have used the phonon Raman spectrum of the diamond from the high
pressure culet or stressed region of the diamond. This is less precise and has
some dependence on the diamond geometry\cite{Baer}. Another method is to embed
a grain of diamond in the pressurization medium, but this also has challenges
and limitations\cite{Dubrovinskaia}. An important gauge is the X-ray spectrum
of metal \textquotedblleft markers\textquotedblright\ embedded in the
pressurization medium, but this is only useful at synchrotrons. Finally we
mention that many of the pressure gauges have problems at high temperatures.
The advantages of the nanoshells, which come dispersed in a volatile fluid, is
that they can be painted onto the diamond culet in a thin invisible (to the
eye) layer so they will not blow out and they cover the entire culet flat so
they cannot easily be masked; they should maintain their sensitivity to the
highest pressures; optical spectroscopy is easy to implement with a large
signal-to-noise ratio, compared to fluorescence or Raman scattering, and will
not be masked by fluorescence from the diamonds. We believe that nanoshells
will maintain their sensitivity at high temperatures, but this requires
further study.

The article is organized as follows. In section II we first review existing
theories that can be used to calculate the spectrum and resonance frequencies
of a nanoshell. Next, the pressure dependency is taken into account. In
section III the results of the calculations are shown for the nanoshell
geometry. To overcome some of the problems seen with nanoshells the coated
nanoshell is introduced in section IV. Section V contains the conclusions.

\section{Theory}

The optical properties of nanoshells are usually described by either Mie
theory or the hybridization model. The hybridization model, developed by
Nordlander and co-workers\cite{Hybri1}, offers a fast-to-evaluate analytical
form for the resonance frequencies, but it is only valid for small particles
(smaller than ca.~a tenth of the wavelength). Mie theory\cite{Stratton} is
valid for all particle sizes but this method is far more time consuming for
calculating the peak positions since it does not allow for a closed form
analytical formula for the position of the resonance peak. These resonance
peak positions have to be numerically derived from the absorption or
extinction spectrum.

In this section first both theories will be reviewed. The pressure dependency
is taken into account in the last part of this section.

\subsection{Hybridization theory}

Hybridization theory\cite{Hybri1} is a phenomenological approach in which the
nanoshell is modeled as a combination of a cavity and a metallic sphere. First
the sphere and cavity are studied separately. The free electrons of the sphere
are descibed as an incompressible fluid excited by plasmons and the dielectric
function is assumed to be that of a Drude metal\cite{Kittel}:%
\begin{equation}
\varepsilon_{s}\left(  \omega\right)  =\varepsilon_{b}-\frac{\omega_{pl}^{2}%
}{\omega^{2}}, \label{Drude}%
\end{equation}
where $\varepsilon_{b}$ indicates the background permittivity and $\omega
_{pl}$ the bulk plasma frequency of the metal.

In a second step the resonance frequencies of the cavity and the sphere are
hybridized due to the coupling between the cavity and the sphere modes, very
similar to the hybridization of molecular orbitals. This results in a closed
formula for the resonance frequencies\cite{Hybri2}:%
\begin{align}
\omega_{L,\pm}^{2}  &  =\frac{\omega_{pl}^{2}}{\beta}\left(  \left(
L+1\right)  ^{2}\varepsilon_{m}+2L\left(  L+1\right)  \varepsilon_{b}%
+L^{2}\varepsilon_{c}+L\left(  L+1\right)  \left(  \varepsilon_{m}%
-2\varepsilon_{b}+\varepsilon_{c}\right)  x^{2L+1}\right. \\
&  \pm\left\{  \left[  L^{2}\varepsilon_{c}-\left(  L+1\right)  ^{2}%
\varepsilon_{m}+L\left(  L+1\right)  \left(  \varepsilon_{m}-\varepsilon
_{c}\right)  x^{2L+1}\right]  ^{2}\right. \nonumber\\
&  \left.  \left.  +4L\left(  L+1\right)  \left[  \left(  L+1\right)
\varepsilon_{m}+L\varepsilon_{c}\right]  ^{2}x^{2L+1}\right\}  ^{1/2}\right)
\nonumber
\end{align}
with:%
\begin{align}
\beta &  =2\left\{  \left[  \left(  L+1\right)  \varepsilon_{b}+L\varepsilon
_{c}\right]  \left[  \left(  L+1\right)  \varepsilon_{m}+L\varepsilon
_{b}\right]  \right. \\
&  \left.  +L\left(  L+1\right)  \left(  \varepsilon_{b}-\varepsilon
_{c}\right)  \left(  \varepsilon_{m}-\varepsilon_{b}\right)  x^{2L+1}\right\}
,\nonumber
\end{align}
where $\varepsilon_{c}$, $\varepsilon_{b}$ and $\varepsilon_{m}$ are the
permittivities of the core, the background of the shell (see formula $\left(
\ref{Drude}\right)  $) and the medium respectively, $L$ is the orbital quantum
number and $x=R_{c}/R_{s}$ is the ratio between the core $\left(
R_{c}\right)  $ and shell $\left(  R_{s}\right)  $ radii.

\subsection{Mie theory}

Mie theory is used to solve the Maxwell equations in systems with spherical
symmetry\cite{Stratton}. It allows the calculation of the absorbed, scattered
and total cross section for an incident beam of radiation as a function of
frequency. The nanoshell is modeled as two concentric spherical spheres. For
this case the results originally derived by Mie for a single spherical
interface were worked out by Aden and Kerker\cite{AdenKerker}. The coated
nanoshell, which will be introduced later, is modeled by three concentric
spheres. Extending Mie theory to three concentric spheres, although not
available in literature, does not provide any new problems in comparison with
the two concentric sphere case solved by Aden and Kerker. The final
expressions for the electromagnetic fields in and around the coated nanoshell,
as well as the scattering and absorption cross section, are given in the
appendix. From the resulting total cross section the resonance frequencies can
be determined numerically.

For this method the input parameters are again the sizes and permittivities of
the shells. In analogy with the hybridization model we used the Drude
dielectric function in Mie theory to describe the permittivity of the metal shell.

\subsection{Pressure dependency: equation of state}

Applying a pressure $P$ to a nanoshell will lead to a change of the volume of
the core $V_{c}\left(  P\right)  $ and the shell $V_{s}\left(  P\right)  $.
These changes can be calculated using the Vinet equation of state
(EOS)\cite{VinetEOS}:%
\begin{equation}
P=3K_{0}\frac{1-\left(  \frac{V}{V_{0}}\right)  ^{1/3}}{\left(  \frac{V}%
{V_{0}}\right)  ^{2/3}}\exp\left\{  \frac{3}{2}\left(  K_{1}-1\right)  \left[
1-\left(  \frac{V}{V_{0}}\right)  ^{1/3}\right]  \right\}  , \label{Vinet EOS}%
\end{equation}
which gives the relation between the pressure $P$ and the relative volume
change $V/V_{0}$ with respect to a reference volume $V_{0}$ of a material at a
certain temperature. Here $K_{0}$ and $K_{1}$ are material constants for which
the values used in this article are presented in table \ref{Table K0 K1}.
\begin{table}[tbp] \centering
\caption{The material properties used in the pressure calculations of the nanoshell. Parameters $K_0$ and $K_1$
are valid up to $580$ GPa for gold and $150$ GPa for amorphous silicon according to resp. ref. \cite{K0K1enV0Au} and ref. \cite{K0K1enV0SiO2}.
The last column indicates the relative permittivity at ambient pressure. The permittivity indicated for gold is
the bulk permittivity $\varepsilon_{b}$ as defined in equation \ref{Drude}.}%
\begin{tabular}
[c]{|l|c|c|c|}\hline
\textbf{Material} & $\mathbf{K}_{0}$ $\left(  \text{GPa}\right)  $ &
$\mathbf{K}_{1}$ & $\varepsilon$\\\hline\hline
Amorphous SiO$_{2}$ & $329$\textsuperscript{\cite{K0K1enV0SiO2}} &
$4.1$\textsuperscript{\cite{K0K1enV0SiO2}} & $2.0449$%
\textsuperscript{\cite{PermSIO2}}\\
Au & $167$\textsuperscript{\cite{K0K1enV0Au}} & $5.94$%
\textsuperscript{\cite{K0K1enV0Au}} & $6.9$%
\textsuperscript{\cite{DrudeBronbackground}}\\\hline
\end{tabular}
\label{Table K0 K1}%
\end{table}
Using the Vinet EOS one can calculate the size change of the nanoshell under
pressure where $V_{0}$ has been taken as the original volume at zero pressure.
Under pressure the nanoshell will be compressed resulting in a different core
and shell radius and a different bulk plasma frequency. These changes can be
calculated from:%
\begin{align}
R_{c}\left(  P\right)   &  =R_{c}\left(  0\right)  \left(  \frac{V_{c}\left(
P\right)  }{V_{c}\left(  0\right)  }\right)  ^{1/3},\\
R_{s}\left(  P\right)   &  =\left\{  R_{s}^{3}\left(  0\right)  \left(
\frac{V_{s}\left(  P\right)  }{V_{s}\left(  0\right)  }\right)  +R_{c}%
^{3}\left(  0\right)  \left[  \left(  \frac{V_{c}\left(  P\right)  }%
{V_{c}\left(  0\right)  }\right)  -\left(  \frac{V_{s}\left(  P\right)
}{V_{s}\left(  0\right)  }\right)  \right]  \right\}  ^{1/3},\\
\omega_{pl}^{2}\left(  P\right)   &  =\frac{n\left(  P\right)  e^{2}%
}{\varepsilon_{0}m}=\left(  \frac{V_{s}\left(  0\right)  }{V_{s}\left(
P\right)  }\right)  \omega_{pl}^{2}\left(  0\right)  , \label{wBP}%
\end{align}%
\begin{figure}[ptb]%
\centering
\includegraphics[
height=6.5833cm,
width=8.6042cm
]%
{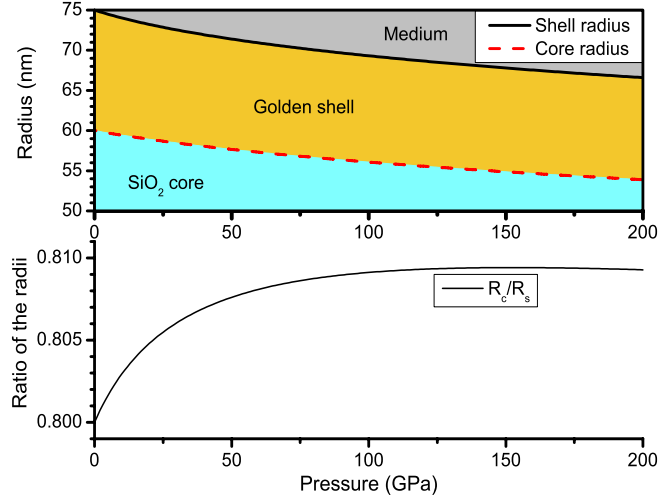}%
\caption{The behavior of the sizes of the nanoshell under pressure. The top
panel presents the radii of the core and shell as a function of pressure.
Notice that the golden shell will be compressed more than the core until about
$150$ GPa where the core/shell ratio reaches a maximum. The bottom panel shows
this ratio $R_{c}/R_{s}$, which is an important parameter for the optical
response. The radii of the nanoshells in this article correspond to the size
of commercially available nanoshells.}%
\label{fig NS under pressure}%
\end{figure}
where $V_{c}\left(  P\right)  $ and $V_{s}\left(  P\right)  $ are the volumes
of the core and the shell respectively, $n$ is the bulk electron
concentration, $e$ is the electron charge, $\varepsilon_{0}$ is the
permittivity of vacuum and $m$ is the electron mass. The pressure dependency
of the radii is shown in Figure \ref{fig NS under pressure} for a nanoshell
with $R_{c}=60%
\operatorname{nm}%
$ and $R_{s}=75%
\operatorname{nm}%
$. When the nanoshell is compressed the electron density will increase and so
will the bulk plasma frequency (\ref{wBP}). This will also influence the
permittivity of the metal since the plasma frequency is an essential part of
the Drude dielectric function (see Eq. $\left(  \ref{Drude}\right)  $).

\begin{figure}[ptb]%
\centering
\includegraphics[
height=6.7649cm,
width=8.6037cm
]%
{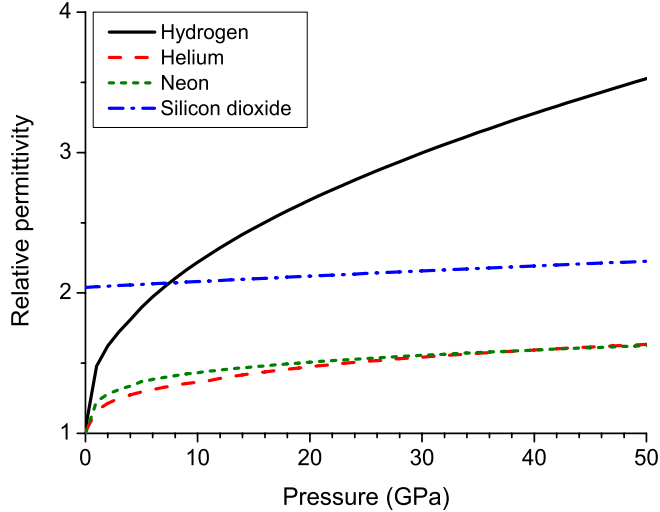}%
\caption{The pressure dependent permittivity of different materials. The
permittivity of SiO$_{2}$ was calculated from the Vinet EOS and the
Clausius-Mossotti relation. The other permittivites are based on
literature\cite{MediaProp}.}%
\label{fig permittivity}%
\end{figure}

The pressure and frequency dependency of the permittivity of SiO$_{2}$ in the
megabar regime remains, to the best of our knowledge, unknown. However it is
possible to estimate the pressure dependency from the Vinet EOS using the
Clausius-Mossotti relation\cite{Kittel}:%
\begin{equation}
\frac{\varepsilon\left(  \omega,P\right)  -1}{\varepsilon\left(
\omega,P\right)  +2}=\frac{4\pi}{3}n\left(  P\right)  \alpha,
\label{Clausius-Mossotti}%
\end{equation}
where $n\left(  P\right)  =N/V\left(  P\right)  $ is the electron density and
$\alpha$ the average atomic polarizability, assumed independent of pressure.
This relation links the permittivity to the volume of a material. Dividing
this equation with the same equation at zero pressure allows us to estimate%
\begin{equation}
\varepsilon(\omega,P)=\frac{2\frac{\varepsilon(\omega,0)-1}{\varepsilon
(\omega,0)+2}+\frac{V_{c}\left(  P\right)  }{V_{c}\left(  0\right)  }}%
{\frac{V_{c}\left(  P\right)  }{V_{c}\left(  0\right)  }-\frac{\varepsilon
(\omega,0)-1}{\varepsilon(\omega,0)+2}}, \label{permittivity from CM}%
\end{equation}
where $V\left(  P\right)  /V_{0}$ can be calculated from the Vinet EOS
$\left(  \ref{Vinet EOS}\right)  $ and $\varepsilon\left(  \omega,0\right)  $
represents the relative permittivity at zero pressure. The pressure dependency
of the permittivity of SiO$_{2}$ and some other possible medium materials is
shown in Figure \ref{fig permittivity}.
Notice that there is no phase change in the SiO$_{2}$ curve. In nature
SiO$_{2}$ is in a crystalline phase at zero pressure and will switch to the
amorphous phase at about $90$ GPa. However, it has been
reported\cite{AmorfeSIO2kern} that the fabrication method of the SiO$_{2}$
particles results in cores which are in the amorphous state also at zero pressure.

\section{Results for a nanoshell}%

\begin{figure}[tbh]%
\centering
\includegraphics[
height=6.502cm,
width=8.6042cm
]%
{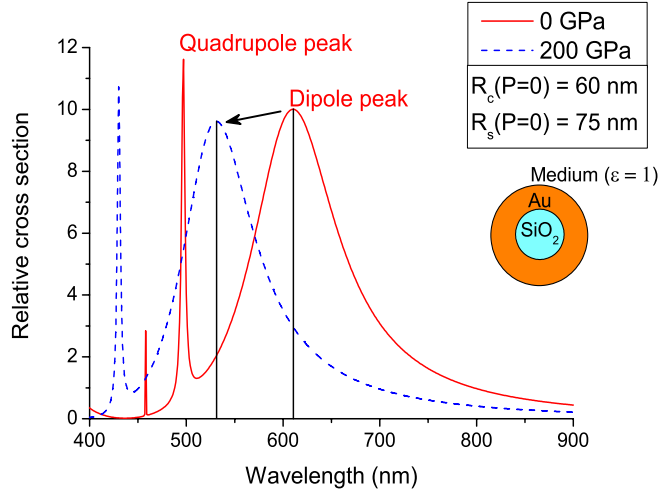}%
\caption{The relative cross section as a function of wavelength for different
pressures for a nanoshell with parameters as mentioned in the figure. Notice
the pressure-induced blueshift of the dipole peak from $611\operatorname{nm}$
at $0$ GPa to $531\operatorname{nm}$\ at $200$ GPa as indicated by the two
black lines.}%
\label{fig rel cross section}%
\end{figure}
Figure \ref{fig rel cross section} shows the relative cross section of the
considered nanoshell placed in vacuum as calculated with Mie theory. The
relative cross section is the total optical cross section divided by $\pi
R_{s}^{2}$, the area of the projection of the nanoshell on a plane
perpendicular to the incoming radiation. From this figure one can clearly see
a substantial blueshift when $200$ GPa pressure is applied. Another important
observation concerns the width of the peak. This is important for increasing
the precision in an experimental measurement since sharper peaks allow for a
more accurate determination of the peak position. However, sharp resonance
lines may elude experimental detection if they do not carry enough spectral
weight. In Figure \ref{fig rel cross section} one can see that the dipole peak
(the rightmost peak) is a broad, clear and rather symmetric peak, while the
quadrupole peak is much sharper and a good candidate for these kind of
experiments. Numerically however it is easier to track the broad dipole peak,
thus all results presented here will be with regard to the dipole peak. It is
seen that although the quantitative results differ for all peaks, the
qualitative results presented here hold true for all resonance peaks.%

\begin{figure}[ptb]%
\centering
\includegraphics[
height=6.4493cm,
width=8.602cm
]%
{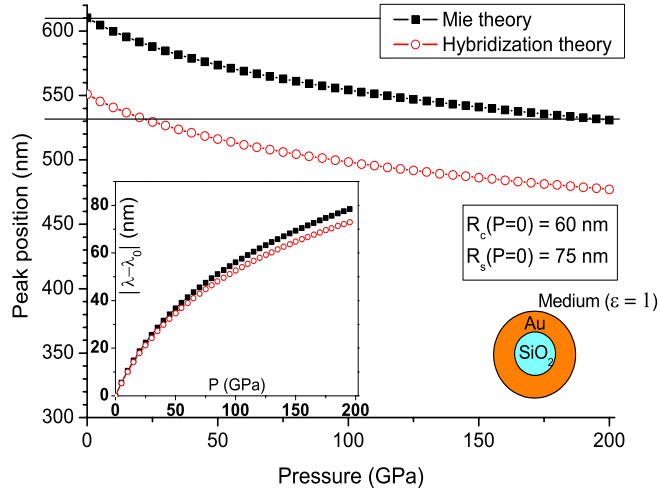}%
\caption{The position of the dipole resonance peak as a function of the
pressure for a nanoshell with parameters as mentioned on the figure. The two
horizontal black lines correspond to the two black lines on figure
\ref{fig rel cross section}. The blueshift from $611\operatorname{nm}$ at $0$
GPa to $531\operatorname{nm}$\ at $200$ GPa as predicted by Mie theory is
clearly visible. The red circles were calculated with hybridization theory and
also indicate a blueshift, but predict a different peak position. The inset
shows the peak shift as a function of pressure compared to the original
position at zero pressure.}%
\label{fig peak position VAC}%
\end{figure}
In Figure \ref{fig peak position VAC} the position of the dipole resonance
peak is shown as a function of pressure for a nanoshell in a medium with
$\varepsilon=1$ constant. The black squares are calculated using Mie theory
and a subsequent numerical determination of the peak maximum. The results from
the hybridization theory are presented by the red circles. It is clear that
the two theories do not agree on the exact position of the resonance peak for
the considered nanoshell. This is due to the electrostatic limit used in the
hybridization model which is only acceptable for nanoparticles much smaller
than the wavelengths of the incident light (the limiting case for nanoshells
much smaller than the wavelengths used in a DAC experiment, with sizes at the
moment unachievable for fabrication, do agree). Still the hybridization model
can prove to be useful because both theories agree well on the amount the peak
shifts, as is shown in the inset of Figure \ref{fig peak position VAC}. The
two theories diverge from each other only at higher pressures. The results
show that the nanoshell could be used to measure the pressure by determing the
amount the resonance peak shifts. The inset is also an indication of the
resolution that could be achieved. For this example nanoshell the dipole
resonance peak shifts over $75%
\operatorname{nm}%
$ when $200$ GPa pressure is applied (for rubies this would be $51.1%
\operatorname{nm}%
$\cite{Ruby error,Rubyto150GPa}).

\begin{figure}[ptb]%
\centering
\includegraphics[
height=8.0572cm,
width=8.6042cm
]%
{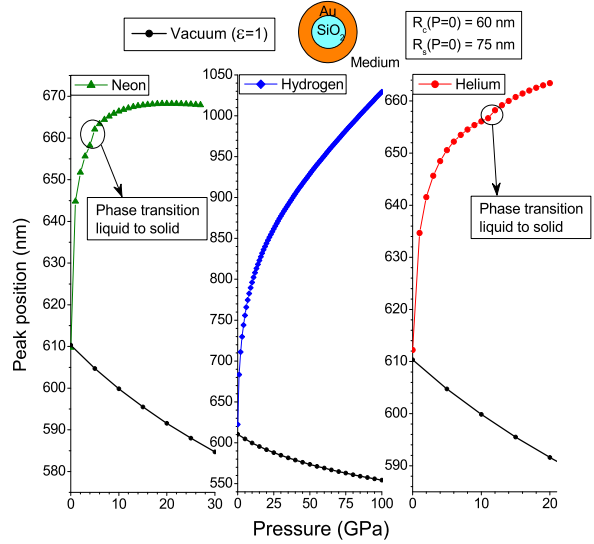}%
\caption{The position of the dipole peak as calculated with Mie theory for
various pressure media: helium\cite{MediaProp} (red circles),
hydrogen\cite{HydrogenSilvera} (blue diamonds) and neon\cite{MediaProp} (green
triangles). In all figures the vacuum position is indicated in black for
reference and the phase transition from liquid to solid is indicated. The
plots only show the pressures for which experimental data is available. For
completeness the original data are presented in table \ref{Table n medium}
together with the pressure range in which they are valid.}%
\label{fig peak position medium}%
\end{figure}

To determine the usefulness of nanoshells as a pressure gauge, calibration
would need to be done for every pressure medium separately. The nanoshell's
optical response depends on the dielectric function of the surrounding medium
in which it is placed, and as such both the position of the resonance peaks
and their pressure-induced shift will differ for different pressure media. The
original data used for the calculation of the dielectric function are shown in
table \ref{Table n medium}. In Figure \ref{fig peak position medium}\
one can see the pressure dependency of the position of the dipole resonance
peak for different pressure media: helium, hydrogen and neon. The behavior of
the medium under pressure clearly has a large influence on the optical
response of the nanoshell. For these media, the blueshift that occurs in
vacuum has turned into a redshift as pressure is increased. From a certain
pressure on, there will be almost no shift of the resonance peak, rendering
the nanoshells ineffectual as pressure gauges in that region. Although it is
possible to calculate the optical response of nanoshells for much higher
pressures, the results are not presented here because they are based on
extrapolation of data from a limited pressure region and as such are
considered unreliable.%

\begin{table}[tbp] \centering
\caption{The refractive index of materials used as pressure medium in a DAC as reported by references \cite{MediaProp} and \cite{HydrogenSilvera}. The last column presents the pressure range of the data on which these fits are based. The dielectric function $\varepsilon$ can be calculated by squaring the refractive index. }%
\begin{tabular}
[c]{|l|c|c|}\hline
\textbf{Material} & \textbf{Refractive index} & $P$\textbf{ range
(GPa)}\\\hline\hline
He (fluid)\textsuperscript{\cite{MediaProp}} & $n=0.8034+0.20256\left(
1+P\right)  ^{0.12763}$ & $0.08-11.5$\\\hline
He (solid)\textsuperscript{\cite{MediaProp}} & $n=-0.1033+\left(  1+P\right)
^{0.052}$ & $11.7-20.2$\\\hline
H$_{2}$\textsuperscript{\cite{HydrogenSilvera}} & $%
\begin{array}
[c]{c}%
n=-0.687343+0.00407826P\\
+1.86605\left(  0.29605+P\right)  ^{0.0646222}%
\end{array}
$ & $0-100$\\\hline
\multicolumn{1}{|l|}{Ne (fluid)\textsuperscript{\cite{MediaProp}}} &
$n=0.668+0.33\left(  1+4.3P\right)  ^{0.076}$ & $0.7-4.7$\\\hline
\multicolumn{1}{|l|}{Ne (solid)\textsuperscript{\cite{MediaProp}}} &
$n=0.9860+0.08597P^{0.1953}$ & $5-27$\\\hline
\end{tabular}
\label{Table n medium}%
\end{table}%

\section{Results for a coated nanoshell}

As was seen in the previous section, the simple nanoshell geometry is not
ideal for high-pressure experiments. The peak shift is affected by the
pressure medium so calibration will be necessary for each nanoshell and
medium. A possible solution would be to shield the nanoparticle from the
effects of the medium, therefore allowing the shift of the resonance peak to
be influenced only by the compression of the nanoparticle. The absence of
these problems when the nanoshell is placed in vacuum indicates that shielding
the particle could indeed result in the desired effect.

Shielding the nanoparticle from the environments means creating a barrier
between the golden shell and the dielectric medium. This barrier should also
be a dielectric. If not, another set of surface plasmons polaritons would
arise on the interface between the outer layer and the environment,
counteracting the intent of the extra layer. In this article the extra coating
is achieved by adding an extra SiO$_{2}$ layer to the model and extending Mie
theory to three concentric spheres. Henceforth we shall indicate this type of
"nanomatryushka"\cite{nanomatr} as "coated nanoshell". The main question to be
answered is how thick this coating should be to effectively shield the
nanoshell from the pressure effects on the dielectric function of the
environment.
\begin{figure}[ptb]%
\centering
\includegraphics[
height=6.2946cm,
width=8.6037cm
]%
{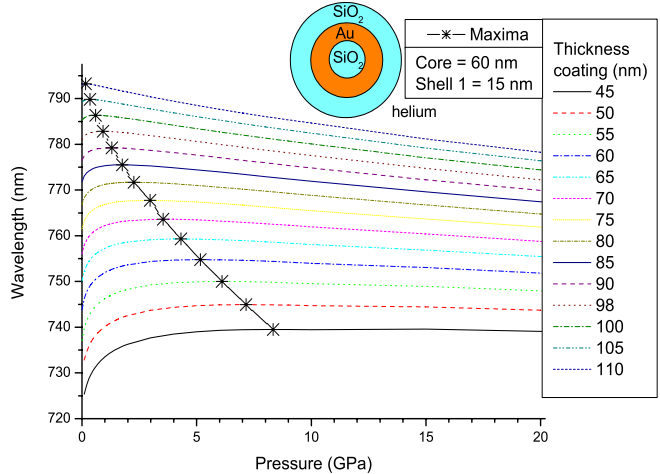}%
\caption{The dipole peak position as a function of pressure for the coated
nanoshell with a $60\operatorname{nm}$ SiO$_{2}$ core, a $15\operatorname{nm}$
thick golden shell and for different thicknesses of the outer SiO$_{2}$
coating. For small coatings the redshift due to the medium is still clearly
visible, while for thick coatings this effect seems to disappear. The black
crosses indicate the pressure at which the wavelength peak position is maximal
and where the redshift turns into a blueshift.}%
\label{fig peak pos CNS}%
\end{figure}
Figure \ref{fig peak pos CNS}\ presents the dipole peak position for several
coating thicknesses of the coated nanoshell as calculated by Mie theory. For
reasons of comparison the core and the thickness of the golden shell are kept
the same as for the nanoshell discussed before. Furthermore we have only
studied the coated nanoshell in a helium environment since this is a commonly
used quasi-hydrostatic pressure medium in DAC experiments. In neon the results
will be almost the same because the dielectric functions of helium and neon
are similar. For hydrogen the counteracting effect is not present, therefore
the results with nanoshells are adequate and no further improvements are necessary.

In this article we do not compare the calculations for the coated nanoshell
with calculations using the hybridization model. The expansion of the
hybridization theory for multiple layered nanoshells is available in
literature\cite{Hybri2}, but it is not considered here due to the expected
deviations since the diameters of the coated nanoshell particles are $2$ or
$3$ times larger than the nanoshell particles and thus the electrostatic
approximation is certainly not valid.

It can be seen from Figure \ref{fig peak pos CNS} that for thin coatings the
redshift due to the effect of the medium is still visible. At a certain
pressure the peak position will reach a maximum and from there on the peak
will undergo a blueshift which is due to the compression of the nanoshell
similar to the results in vacuum (Figure \ref{fig peak position VAC}). This
maximum is indicated by the black crosses and can be used as a measure for the
pressure up to which the medium affects the optical response of the
nanoparticle. It is clear that for thicker coatings the maximum shifts to
lower pressures suggesting that the influence of the medium indeed diminishes
and thus that the coating effectively shields the nanoparticle. Unfortunately
the redshift never disappears completely meaning that the influence of the
medium cannot be fully shielded by the dielectric layer. However, it is
possible to position the maximum into a pressure region where it can do no
harm for the pressure measurements.

Obviously, the thickness of the coating has to be adjusted to the needs of the
experiments. A thicker coating will provide better shielding, but bigger
particles take up more space and risk interacting with each other due to being
in close proximity of one another. However, since the coated nanoshell is much
larger than the nanoshell, the absolute cross section is larger, meaning that
a smaller density is sufficient while still gaining the same response.

The derivative of the curves on Figure \ref{fig peak pos CNS} give a direct
indication of the sensitivity achievable with the coated nanoshell structures.
On average the sensitivity for the coated nanoshell with a $110%
\operatorname{nm}%
$ thick coating (upper curve on Figure \ref{fig peak pos CNS}) is
approximately $0.90$ $%
\operatorname{nm}%
/$GPa. The resolution for rubies in the same pressure regime ($0$ to $20$ GPa)
is approximately $0.36$ $%
\operatorname{nm}%
/$GPa\cite{Ruby error} and for a nanoshell in vacuum as was shown in Figure
\ref{fig peak position VAC} this would be $0.94$ $%
\operatorname{nm}%
/$GPa. From this we can conclude that the theory predicts a better resolution
for nanoshells than for rubies in the pressure regime under consideration.

\section{Discussion and Conclusions}

In this article we have carried out a theoretical analysis of nanoshells which
can be designed so that they have absorption peaks in the IR-visible part of
the optical spectrum due to scattering or absorption by localized surface
plasmon polaritons. Calculations indicate that for nanoshells in vacuum these
peaks have a substantial shift with pressure making them suitable as a
pressure gauge for high-pressure research. A useful pressure gauge should have
a calibration (peak wavelength vs pressure) independent of the pressurization
medium. We found that for a simple nanoshell consisting of an SiO$_{2}$ core
and a gold shell, the calibration differed with the pressurization medium;
moreover it was double-valued and had a region of zero slope. The latter
problem was resolved by coating the nanoshell with an SiO$_{2}$ cladding,
resulting in a robust sensitive pressure gauge. The proposed nanoshell gauge
has advantages in comparison with well-established pressure gauges: they will
not easily blow out during loading of a DAC, they cannot easily be masked by
fluorescence from the diamonds, they have large signal-to-noise ratio in
comparison to fluorescence and Raman Scattering, and due to their small
dimensions they will not be stressed by bridging between the diamonds at very
high pressure when the gasket thins. All of these putative advantages must be
confirmed by experiment.

A possible implementation of the coated nanoshell would be to distribute them
inside the DAC cell together with the ruby on the diamond culet. For low
pressures both gauges can be used and the coated nanoshell can be calibrated
by using the extensive knowledge of the behavior of ruby under
pressure\cite{Ruby error}. A possible challenge is the application of
nanoshells to the surface of a diamond culet. Coated nanoshells can be
acquired at high concentration $\left(  2\sim3\times10^{9}/%
\operatorname{mm}%
^{3}\right)  $ from Nanospectra Biosciences, Inc., in a liquid solution. A
droplet can be placed on the culet and allowed to evaporate to produce a
coverage bonded to the surface by van der Waals forces. Preliminary
measurements show that to avoid clustering and segregation, it may prove
useful to functionalize the diamond surface with a film of
poly-4-vinylpyridine (\textsc{pvp}) which has dense sites that localize the
nanoparticles\cite{PVP}.\ For high pressures the ruby measurement would be
difficult or no longer be possible. The spectra of the coated nanoshells
however will still be measurable since these measurements are based on
absorption and transmission. In this way nanoshells could be easily used and
effectively extend pressure measurements to ultra high pressures.

\begin{acknowledgments}
We thank No\'{e}mie Bardin for aiding with prelimary measurements of
distributing coated nanoshells on diamond culets. Research funded by a Ph.D.
grant of the Agency for Innovation by Science and Technology (IWT). This work
is supported financially by the Fund for Scientific Research Flanders, FWO
project G.0365.08, and NSF grant DMR-0804378, DoE SSAA grant DE-FG52-10NA29656.
\end{acknowledgments}

\appendix{}

\section{Absorption and scattering of an electromagnetic wave by 3 concentric
spheres.}

\begin{figure}[ptbh]%
\centering
\includegraphics[
height=4.525cm,
width=4.0857cm
]%
{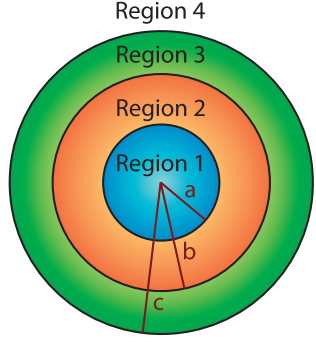}%
\caption{Schematic overview of the modeled system: region $1$ will be the
SiO$_{2}$ core, region $2$ the golden shell, region $3$ the SiO$_{2}$ coating
and region $4$ the pressure medium.}%
\label{fig schema 3 concentric spheres}%
\end{figure}

The coated nanoshell was modeled using Mie theory for 3 concentric spheres.
These spheres divide the space into 4 regions: the core of the nanoparticle,
two shells and the medium in which the nanoparticle is immersed, respectively
numbered $1$ through $4$ as shown in Figure
\ref{fig schema 3 concentric spheres}.
The radii of the three spheres are indicated with $a,b$ and $c$, from inner to
outer. Using the notations of Bohren and Huffman\cite{bohren} the incoming
plane wave can be expanded in vector spherical harmonics:%
\begin{align}
\vec{E}_{4,i}  &  =\sum_{n=1}^{\infty}E_{n}\left(  \vec{M}_{o1n}^{\left(
1\right)  }-i\vec{N}_{e1n}^{\left(  1\right)  }\right)  ,\\
\vec{H}_{4,i}  &  =\frac{-k_{4}}{\omega\mu_{4}}\sum_{n=1}^{\infty}E_{n}\left(
\vec{M}_{e1n}^{\left(  1\right)  }+i\vec{N}_{o1n}^{\left(  1\right)  }\right)
,
\end{align}
with:%
\begin{equation}
E_{n}=E_{0}i^{n}\frac{2n+1}{n\left(  n+1\right)  }.
\end{equation}
The vector spherical harmonics can be written as:%
\begin{align}
\vec{M}_{e1n}^{\left(  i\right)  }  &  =-z_{n}^{\left(  i\right)  }\left(
kr\right)  \pi_{n}\left(  \cos\theta\right)  \sin\varphi\vec{e}_{\theta}%
-z_{n}^{\left(  i\right)  }\left(  kr\right)  \tau_{n}\left(  \cos
\theta\right)  \cos\varphi\vec{e}_{\varphi},\\
\vec{M}_{o1n}^{\left(  i\right)  }  &  =z_{n}^{\left(  i\right)  }\left(
kr\right)  \pi_{n}\left(  \cos\theta\right)  \cos\varphi\vec{e}_{\theta}%
-z_{n}^{\left(  i\right)  }\left(  kr\right)  \tau_{n}\left(  \cos
\theta\right)  \sin\varphi\vec{e}_{\varphi},\\
\vec{N}_{e1n}^{\left(  i\right)  }  &  =\left(  \frac{1}{kr}z_{n}^{\left(
i\right)  }\left(  kr\right)  n\left(  n+1\right)  \pi_{n}\left(  \cos
\theta\right)  \sin\theta\cos\varphi\vec{e}_{r}\right. \\
&  \left.  +dz_{n}^{\left(  i\right)  }\left(  kr\right)  \tau_{n}\left(
\cos\theta\right)  \cos\varphi\vec{e}_{\theta}-dz_{n}^{\left(  i\right)
}\left(  kr\right)  \pi_{n}\left(  \cos\theta\right)  \sin\varphi\vec
{e}_{\varphi}\right)  ,\nonumber\\
\vec{N}_{o1n}^{\left(  i\right)  }  &  =\left(  \frac{1}{kr}z_{n}\left(
kr\right)  n\left(  n+1\right)  \pi_{n}\left(  \cos\theta\right)  \sin
\theta\sin\varphi\vec{e}_{r}\right. \\
&  \left.  +dz_{n}^{\left(  i\right)  }\left(  kr\right)  \tau_{n}\left(
\cos\theta\right)  \sin\varphi\vec{e}_{\theta}+dz_{n}^{\left(  i\right)
}\left(  kr\right)  \pi_{n}\left(  \cos\theta\right)  \cos\varphi\vec
{e}_{\varphi}\right)  ,\nonumber
\end{align}
with the angle dependent functions (where $P_{n}^{m}\left(  x\right)  $ are
the associated Legendre polynomials):%
\begin{align}
\pi_{n}\left(  \cos\theta\right)   &  =\frac{P_{n}^{1}\left(  \cos
\theta\right)  }{\sin\theta},\\
\tau_{n}\left(  \cos\theta\right)   &  =\frac{dP_{n}^{1}\left(  \cos
\theta\right)  }{d\theta},
\end{align}
and the radial dependent functions:%
\begin{align}
z_{n}^{1}\left(  kr\right)   &  =j_{n}\left(  kr\right)  ,\\
z_{n}^{2}\left(  kr\right)   &  =y_{n}\left(  kr\right)  ,\\
z_{n}^{3}\left(  kr\right)   &  =h_{n}^{\left(  1\right)  }\left(  kr\right)
.
\end{align}
On the right hand side one can see the wel known spherical Bessel, Neumann and
Hankel functions respectively.

In region $4$ there will also be a scattered wave which can be described by:%
\begin{align}
\vec{E}_{4,s}  &  =\sum_{n=1}^{\infty}E_{n}\left(  -a_{n}\vec{M}%
_{o1n}^{\left(  3\right)  }+ib_{n}\vec{N}_{e1n}^{\left(  3\right)  }\right)
,\label{eq scatt E}\\
\vec{H}_{4,s}  &  =\frac{k_{4}}{\omega\mu_{4}}\sum_{n=1}^{\infty}E_{n}\left(
b_{n}\vec{M}_{e1n}^{\left(  3\right)  }+ia_{n}\vec{N}_{o1n}^{\left(  3\right)
}\right)  , \label{eq scatt H}%
\end{align}
where $a_{n}$ and $b_{n}$ are coefficients to be determined from the boundary
conditions. The total electromagnetic wave in region $4$ is therefore given by
the sum of the incoming and the scattered fields.

The electromagnetic fields in the other three regions are given by:%

\begin{align}
&  \left\{
\begin{array}
[c]{c}%
\vec{E}_{3}=\sum_{n=1}^{\infty}E_{n}\left(  c_{n}^{\left(  1\right)  }\vec
{M}_{o1n}^{\left(  1\right)  }+c_{n}^{\left(  2\right)  }\vec{M}%
_{o1n}^{\left(  2\right)  }-id_{n}^{\left(  1\right)  }\vec{N}_{e1n}^{\left(
1\right)  }-id_{n}^{\left(  2\right)  }\vec{N}_{e1n}^{\left(  2\right)
}\right) \\
\vec{H}_{3}=\frac{-k}{\omega\mu}\sum_{n=1}^{\infty}E_{n}\left(  d_{n}^{\left(
1\right)  }\vec{M}_{e1n}^{\left(  1\right)  }+d_{n}^{\left(  2\right)  }%
\vec{M}_{e1n}^{\left(  2\right)  }+ic_{n}^{\left(  1\right)  }\vec{N}%
_{o1n}^{\left(  1\right)  }+ic_{n}^{\left(  2\right)  }\vec{N}_{o1n}^{\left(
2\right)  }\right)
\end{array}
\right.  ,\\
&  \left\{
\begin{array}
[c]{c}%
\vec{E}_{2}=\sum_{n=1}^{\infty}E_{n}\left(  e_{n}^{\left(  1\right)  }\vec
{M}_{o1n}^{\left(  1\right)  }+e_{n}^{\left(  2\right)  }\vec{M}%
_{o1n}^{\left(  2\right)  }-if_{n}^{\left(  1\right)  }\vec{N}_{e1n}^{\left(
1\right)  }-if_{n}^{\left(  2\right)  }\vec{N}_{e1n}^{\left(  2\right)
}\right) \\
\vec{H}_{2}=\frac{-k}{\omega\mu}\sum_{n=1}^{\infty}E_{n}\left(  f_{n}^{\left(
1\right)  }\vec{M}_{e1n}^{\left(  1\right)  }+f_{n}^{\left(  2\right)  }%
\vec{M}_{e1n}^{\left(  2\right)  }+ie_{n}^{\left(  1\right)  }\vec{N}%
_{o1n}^{\left(  1\right)  }+ie_{n}^{\left(  2\right)  }\vec{N}_{o1n}^{\left(
2\right)  }\right)
\end{array}
\right.  ,\\
&  \left\{
\begin{array}
[c]{c}%
\vec{E}_{1}=\sum_{n=1}^{\infty}E_{n}\left(  g_{n}\vec{M}_{o1n}^{\left(
1\right)  }-is_{n}\vec{N}_{e1n}^{\left(  1\right)  }\right) \\
\vec{H}_{1}=\frac{-k}{\omega\mu}\sum_{n=1}^{\infty}E_{n}\left(  s_{n}\vec
{M}_{e1n}^{\left(  1\right)  }+ig_{n}\vec{N}_{o1n}^{\left(  1\right)
}\right)
\end{array}
\right.  .
\end{align}
Using the boundary conditions:%
\begin{align}
\vec{E}_{j}\times\vec{e}_{r}  &  =\vec{E}_{j+1}\times\vec{e}_{r},\\
\vec{H}_{j}\times\vec{e}_{r}  &  =\vec{H}_{j+1}\times\vec{e}_{r}%
\end{align}
at each interface, where $\vec{e}_{r}$ indicates the radial unit vector which
is perpendicular to the boundary, we obtain a system of 12 coupled equations
to be solved. This system of equations can be split in two sets of six
equations that we write in matrix form $A_{i}\cdot x_{i}=b_{i}$ with for the
first set of equations%
\begin{equation}
A_{1}=%
\begin{bmatrix}
0 & 0 & 0 & j_{n}\left(  k_{2}a\right)  & y_{n}\left(  k_{2}a\right)  &
-j_{n}\left(  k_{1}a\right) \\
0 & 0 & 0 & dj_{n}\left(  k_{2}a\right)  & dy_{n}\left(  k_{2}a\right)  &
-\frac{\mu_{2}k_{1}}{\mu_{1}k_{2}}dj_{n}\left(  k_{1}a\right) \\
h_{n}^{\left(  1\right)  }\left(  k_{4}c\right)  & j_{n}\left(  k_{3}c\right)
& y_{n}\left(  k_{3}c\right)  & 0 & 0 & 0\\
\frac{k_{4}\mu_{3}}{\mu_{4}k_{3}}dh_{n}^{\left(  1\right)  }\left(
k_{4}c\right)  & dj_{n}\left(  k_{3}c\right)  & dy_{n}\left(  k_{3}c\right)  &
0 & 0 & 0\\
0 & j_{n}\left(  k_{3}b\right)  & y_{n}\left(  k_{3}b\right)  & -j_{n}\left(
k_{2}b\right)  & -y_{n}\left(  k_{2}b\right)  & 0\\
0 & \frac{\mu_{2}k_{3}}{\mu_{3}k_{2}}dj_{n}\left(  k_{3}b\right)  & \frac
{\mu_{2}k_{3}}{\mu_{3}k_{2}}dy_{n}\left(  k_{3}b\right)  & -dj_{n}\left(
k_{2}b\right)  & -dy_{n}\left(  k_{2}b\right)  & 0
\end{bmatrix}
,
\end{equation}
and%
\begin{equation}
b_{1}=%
\begin{bmatrix}
0\\
0\\
j_{n}\left(  k_{4}c\right) \\
\frac{k_{4}\mu_{3}}{\mu_{4}k_{3}}dj_{n}\left(  k_{4}c\right) \\
0\\
0
\end{bmatrix}
,\qquad x_{1}=%
\begin{bmatrix}
a_{n}\\
c_{n}^{\left(  1\right)  }\\
c_{n}^{\left(  2\right)  }\\
e_{n}^{\left(  1\right)  }\\
e_{n}^{\left(  2\right)  }\\
g_{n}%
\end{bmatrix}
.
\end{equation}
And for the second set of equations%
\begin{equation}
A_{2}=%
\begin{bmatrix}
0 & 0 & 0 & dj_{n}\left(  k_{2}a\right)  & dy_{n}\left(  k_{2}a\right)  &
-dj_{n}\left(  k_{1}a\right) \\
0 & 0 & 0 & j_{n}\left(  k_{2}a\right)  & y_{n}\left(  k_{2}a\right)  &
-\frac{\mu_{2}k_{1}}{\mu_{1}k_{2}}j_{n}\left(  k_{1}a\right) \\
dh_{n}^{\left(  1\right)  }\left(  k_{4}c\right)  & dj_{n}\left(
k_{3}c\right)  & dy_{n}\left(  k_{3}c\right)  & 0 & 0 & 0\\
\frac{k_{4}\mu_{3}}{\mu_{4}k_{3}}h_{n}^{\left(  1\right)  }\left(
k_{4}c\right)  & j_{n}\left(  k_{3}c\right)  & y_{n}\left(  k_{3}c\right)  &
0 & 0 & 0\\
0 & dj_{n}\left(  k_{3}b\right)  & dy_{n}\left(  k_{3}b\right)  &
-dj_{n}\left(  k_{2}b\right)  & -dy_{n}\left(  k_{2}b\right)  & 0\\
0 & \frac{\mu_{2}k_{3}}{\mu_{3}k_{2}}j_{n}\left(  k_{3}b\right)  & \frac
{\mu_{2}k_{3}}{\mu_{3}k_{2}}y_{n}\left(  k_{3}b\right)  & -j_{n}\left(
k_{2}b\right)  & -y_{n}\left(  k_{2}b\right)  & 0
\end{bmatrix}
,
\end{equation}%
\begin{equation}
b_{2}=%
\begin{bmatrix}
0\\
0\\
dj_{n}\left(  k_{4}c\right) \\
\frac{k_{4}\mu_{3}}{\mu_{4}k_{3}}j_{n}\left(  k_{4}c\right) \\
0\\
0
\end{bmatrix}
,\qquad x_{2}=%
\begin{bmatrix}
b_{n}\\
d_{n}^{\left(  1\right)  }\\
d_{n}^{\left(  2\right)  }\\
f_{n}^{\left(  1\right)  }\\
f_{n}^{\left(  2\right)  }\\
s_{n}%
\end{bmatrix}
.
\end{equation}
These equations are solved by matrix inversion.\ Hence the total solution in
every region is known. The optical cross section can now be calculated by:%
\begin{align}
Q_{scatt}  &  =\frac{2\pi}{k_{4}^{2}}\sum_{n=1}^{\infty}\left(  2n+1\right)
\left(  \left\vert a_{n}\right\vert ^{2}+\left\vert b_{n}\right\vert
^{2}\right)  ,\\
Q_{abs}  &  =\frac{2\pi}{k_{4}^{2}}\sum_{n=1}^{\infty}\left(  2n+1\right)
\left(  \left\vert a_{n}\right\vert ^{2}+\left\vert b_{n}\right\vert
^{2}-\operatorname{Re}\left(  a_{n}+b_{n}\right)  \right)  ,\\
Q_{tot}  &  =\frac{2\pi}{k_{4}^{2}}\sum_{n=1}^{\infty}\left(  2n+1\right)
\operatorname{Re}\left(  a_{n}+b_{n}\right)  ,
\end{align}
where $a_{n}$ and $b_{n}$ are the coefficients of the scattered wave as
introduced in equations $\left(  \ref{eq scatt E}\right)  $ and $\left(
\ref{eq scatt H}\right)  $. The wave vector $k$ can be calculated from
\begin{equation}
k_{i}^{2}=\omega^{2}\varepsilon_{i}\mu_{i}%
\end{equation}
with $\varepsilon_{i}$ de permittivity and $\mu_{i}$ the permeability of the
$i^{the}$ region.


\bigskip

\begin{thebibliography}{99}                                                                                               %


\bibitem {NS}L.R. Hirsch, A.M. Gobin, A.R. Lowery, F. Tam, R.A. Drezek, N.J.
Halas and J.L. West, Annals of biomedical engineering \textbf{34}, 15 $\left(
2006\right)  $.

\bibitem {cancer}J. Yang, J.\ Lee, J. Kang, S.J. Oh, H.-J. Ko, J.-H. Son, K.
Lee, J-S.. Suh, Y.-M. Huh and S. Haam, Adv. Matter \textbf{21}, 1 $\left(
2009\right)  $.

\bibitem {biosensor}M.L. Brongersma, Nature materials \textbf{2}, 296 $\left(
2003\right)  $.

\bibitem {sers}J.B. Jackson, S.L. Westcott, L.R. Hirsch, J.L. West and N.J.
Halas, Appl. Phys. Letters \textbf{82}, 257 $\left(  2003\right)  $.

\bibitem {Plasmonics}S. Lal, S. Link and N.J. Halas, Nature photonics
\textbf{1}, 641 $\left(  2007\right)  $.

\bibitem {DAC}A. Jayaraman, Rev. Mod. Phys \textbf{55}, 1 $\left(
1983\right)  $.

\bibitem {Ruby}H.K. Mao, J. Xu and P.M. Bell, Journal of geophysical research
\textbf{91,} 4673 $\left(  1986\right)  $.

\bibitem {Rubyscale}G.J. Piermarini, S.J. Block, J.D. Barnett and R.A. Forman,
J. Appl. Phys., \textbf{46,} 2774 $\left(  1975\right)  $.

\bibitem {Ruby error}I.F. Silvera, A.D. Chijioke, W.J. Nellis, A. Soldatov and
J. Tempere, Phys. Stat. Sol (b) \textbf{244}, 460 $\left(  2007\right)  $.

\bibitem {Rubyto150GPa}A. Chijioke, W.J. Nellis, A. Soldatov and I.F. Silvera,
J. Appl. Phys. \textbf{98}, 114905, $\left(  2005\right)  .$

\bibitem {Eggert}J.H. Eggert, K.A. Goetel and I.F. Silvera, Appl. Phys.
Letters \textbf{53}, 2489 $\left(  1988\right)  .$

\bibitem {Chen}N.H. Chen and I.F. Silvera, Rev. Sci. Instrum. \textbf{67},
4275 $\left(  1996\right)  .$

\bibitem {Baer}B.J. Baer, M.E., Chang and W.J. Evans, J. Appl. Phys.
\textbf{104}, 034504 $\left(  2008\right)  $.

\bibitem {Dubrovinskaia}N. Dubrovinskaia, L. Dubrovinsky, R. Caracas and M.
Hanfland, Appl. Phys. Letters \textbf{97}, 251903 $\left(  2010\right)  .$

\bibitem {Hybri1}E. Prodan, C. Radloff, N.J. Halas and P. Nordlander, Science
\textbf{302}, 419 $\left(  2003\right)  $.

\bibitem {Stratton}J.A. Stratton, \textit{Electromagnetic theory} (McGraw-Hill
book company, New York and London $1941$), p.563-573.

\bibitem {Kittel}C. Kittel, \textit{Introduction to solid state physics, fifth
edition} (John Wiley \& Sons, USA $1976$), p.410.

\bibitem {nanomatr}C. Radloff and N.J. Halas, Nano Letters \textbf{4}, 1323
$\left(  2004\right)  $.

\bibitem {Hybri2}E. Prodan and P. Nordlander, J. Chem. Phys. \textbf{120},
5444 $\left(  2004\right)  $.

\bibitem {AdenKerker}A.L. Aden and M. Kerker, J. Appl. Phys. \textbf{22}, 1242
$\left(  1951\right)  $.

\bibitem {VinetEOS}P. Vinet, J. Ferrante, J.H. Rose and J.R. Smith, Journal of
geophysical research \textbf{92}, 9319 $\left(  1987\right)  $.

\bibitem {K0K1enV0SiO2}K.P. Driver, R.E. Cohen, Z. Wu, B. Militzer, P.L. Rios,
M.D.\ Towler, R.J. Needs and J.W. Wilkins, Proc. Natl. Acad. Sci. USA
\textbf{107}, 9519 $\left(  2010\right)  $.

\bibitem {PermSIO2}C.E. Rayford II, G. Schatz and K. Shuford, Nanoscape
\textbf{2}, 27 $\left(  2005\right)  $.

\bibitem {K0K1enV0Au}M. Yokoo, N. Kawai, K.G. Nakamura, K-I Kondo, Y. Tange
and T. Tsuchiya, Phys. Rev. B \textbf{80}, 104114 $\left(  2009\right)  $.

\bibitem {DrudeBronbackground}I.N. Shklyarefskii and P.L. Pakhmov, USSR optika
i Spektroskopiya \textbf{34}, 163 $\left(  1973\right)  $ also at
http://www.mit.edu/\symbol{126}6.777/matprops/gold.htm.

\bibitem {AmorfeSIO2kern}S. Kalele, S.W. Gosavi, J. Urban and S.K. Kulkarni,
Current science \textbf{19}, 1038 $\left(  2006\right)  $.

\bibitem {MediaProp}A. Dewaele, J.H. Egert, P. Loubeyre and R. Le Toullec,
Phys. Rev. B \textbf{67}, 094112 $\left(  2003\right)  $.

\bibitem {HydrogenSilvera}W.J. Evans and I. F. Silvera, Phys. Rev. B
\textbf{57}, 14105 $\left(  1998\right)  $.

\bibitem {bohren}C.F. Bohren and D.R. Huffman, \textit{Absorption and
scattering of light by small particles }(Wiley-VCH, Germany $2004$), p.82-104.

\bibitem {PVP}S. Malynych, I. Luzinov and G. Chymanov, J. Phys. Chem. B
\textbf{106}, 1280 $\left(  2002\right)  $.
\end{thebibliography}
\end{document}